\documentstyle[twoside,fleqn,espcrc2,epsf]{article}

\def\dirac{{\rm D}\!\!\!\!/\,}
\def\lpmb#1{\mbox{\boldmath$#1$}}

\title{
\vspace{-0.5cm}
Predicting the Aoki Phase using the Chiral Lagrangian%
\thanks{Supported by DOE
contract DE-FG03-96ER40956.}
}

\author{
Stephen Sharpe$^{a}$
and 
Robert L. Singleton Jr.\address{Department of Physics,
University of Washington, Seattle, WA 98195, USA}\thanks{Speaker.}
}
       
\begin{document}

\begin{abstract}
This work is concerned with the phase diagram of Wilson fermions 
in the mass and coupling constant plane for two-flavor (unquenched) QCD. We 
show that as the continuum limit is approached, one can study the 
lattice theory using the continuum chiral Lagrangian, supplemented 
by additional terms proportional to powers of the lattice spacing. 
We find two possible phase structures at non-zero
lattice spacing: (1) There is an Aoki phase of spontaneously broken
flavor and parity, with two massless Goldstone-pions,
and a width $\Delta m_0 \sim a^3$; (2) There is no spontaneous symmetry
breaking, and all three pions have equal mass of order $a$. 
Present numerical simulations suggest that the former option is 
realized. 
\end{abstract}

\maketitle

\begin{figure}[t]
\vskip0.1cm
\epsfxsize=3.0in
\centerline{\epsffile{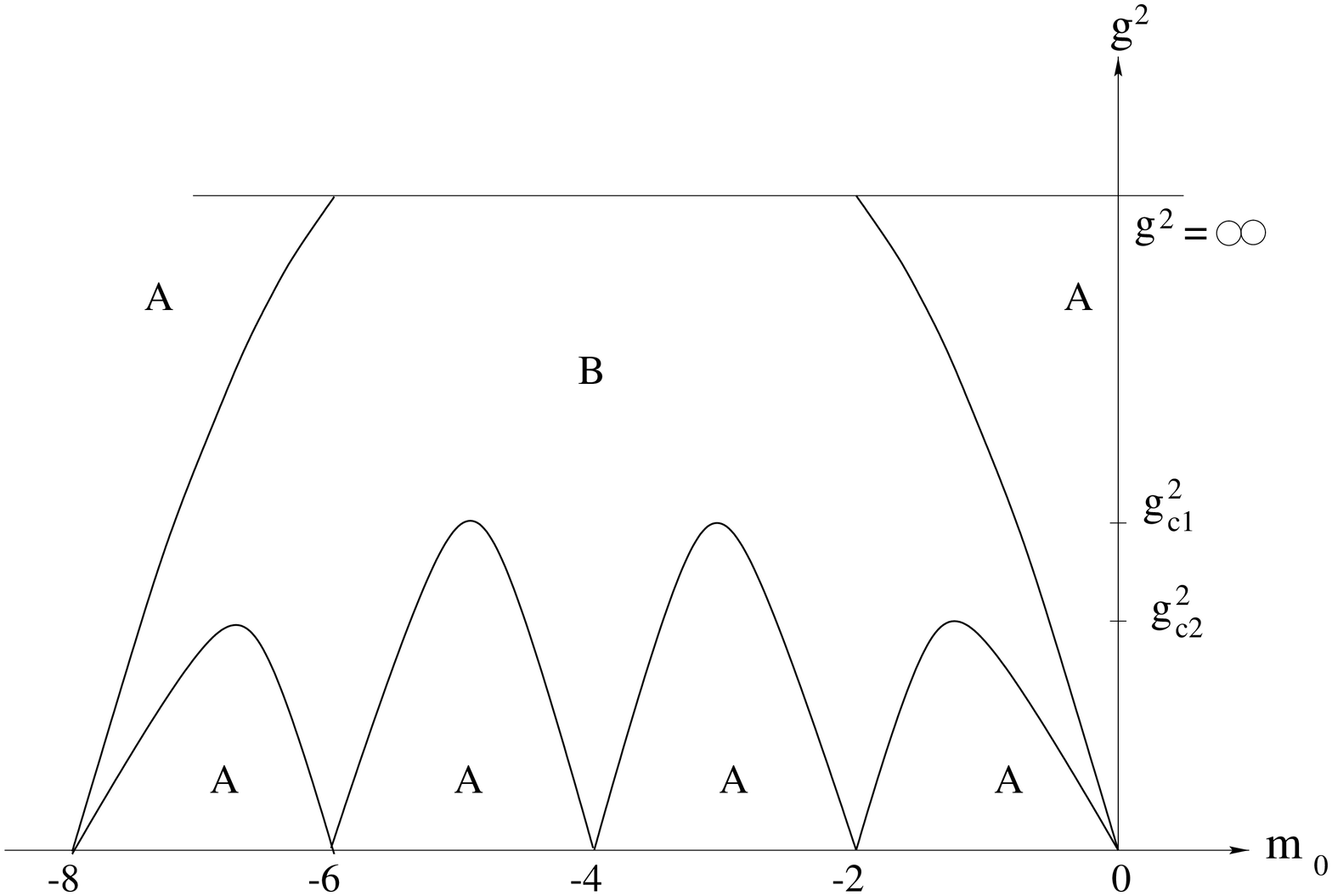}}
\vspace{-0.2truein}
\caption{
The phase diagram proposed by Aoki: $g$ is the gauge
coupling and $m_0$ the dimensionless bare quark-mass. 
The continuum-like unbroken phases are labeled $A$, and 
the flavor and parity broken phase $B$. The continuum 
limit of particular interest is that at $m_0=0$, $g=0$.
}
\vspace{-0.2truein}
\label{fig:aoki}
\end{figure}

Some time ago, Aoki proposed that lattice QCD with Wilson fermions 
has a phase in which flavor and parity are spontaneously broken,
thereby providing a dynamical explanation for the presence of 
massless pions at non-zero lattice spacing\cite{aokiphase}.
His proposed phase diagram is shown in Fig.~\ref{fig:aoki}.
This suggestion has been studied using analytical and numerical methods,
and somewhat contradictory conclusions have been drawn\footnote{For
detailed references to earlier work see Ref. \cite{ss}}.
We have recently shown that one can analyze the phase structure
analytically close to the continuum limit\cite{ss}.
In this paper we summarize
the salient points of our work\footnote{M.~Creutz had 
previously performed a similar analysis based 
on the linear sigma model\cite{creutz}.}.

We consider lattice QCD with two degenerate Wilson fermions.
The action is invariant under flavor $SU(2)$ and parity, but explicitly
breaks all axial symmetries for any value of the bare mass. 
Aoki's proposal is that the condensate $\langle \bar\psi i \gamma_5 
\sigma_3 \psi \rangle$ is non-zero in region $B$ (with
$\sigma_3$ acting in flavor space), thus breaking the
flavor symmetry to a $U(1)$ subgroup and producing two 
massless Goldstone pions throughout $B$. 

To investigate this suggestion analytically, we first write 
down an effective continuum description of the lattice theory valid 
close to the continuum limit, and then examine the long 
distance behavior of the effective theory using standard chiral 
Lagrangian techniques. The effective continuum theory can be 
described by a Lagrangian in which the usual terms have been 
supplemented by higher dimensional operators proportional to 
powers of the lattice spacing\cite{symanzik}. These additional 
pieces are constrained only by the symmetries of the lattice 
action, which means that they need not respect chiral symmetry, 
and their enumeration is identical to that carried out as part 
of the improvement program\cite{luscher}. As will become apparent
shortly, we are interested in physical masses of order the 
lattice spacing, {\em i.e.} $m \sim a \Lambda^2$, with $\Lambda$ an
abbreviation for $\Lambda_{\rm QCD}$. 
In this mass regime, with a suitable choice 
of quark variables, the effective continuum description of the 
lattice theory can be written\cite{ss}
\begin{eqnarray}
\label{leff2}
  {\cal L}_{\rm eff} = {\cal L}_{\rm g} + 
  \bar\psi (\dirac+ m ) \psi +
  b_1 a \bar\psi i \sigma_{\mu\nu} F_{\mu\nu}\psi \ ,
\end{eqnarray}
where ${\cal L}_{\rm g}$ is the gluon Lagrangian,
and we have dropped terms of ${\cal O}(a^2)$. 
The dimensionless parameter $b_1\sim 1$ 
is a function of the coupling $g$,
while the physical mass is $m\sim (m_0-\tilde m_c(g))/a$,
with $m_0$ the (dimensionless) bare mass and $\tilde m_c$ a critical 
mass.\footnote{The symbol $\sim$ means that we are not attempting
to keep track of renormalization factors of order unity.}
Both the mass term and the Pauli term in (\ref{leff2}) break chiral
symmetry. A key point in the following is that when we adjust
the bare mass at fixed $g$ (and thus fixed $a$)
we change the relative importance of these two terms.

The next step is to write a generalization of the
continuum chiral Lagrangian that includes the
effects of the Pauli term. Without either the mass
term or the Pauli term, the theory is invariant
under $SU(2)_L \times SU(2)_R$ chiral rotations,
and its low momentum dynamics is described by the
chiral Lagrangian
\begin{equation}
 {\cal L}_\chi = {f_\pi^2\over 4} \,
  {\sf Tr}\, \left(\partial^\mu\Sigma^\dagger 
  \partial_\mu\Sigma \right) \,.
\end{equation}
Here $\Sigma$ is an $SU(2)$ matrix-valued field that 
transforms under the chiral group as 
$\Sigma \to L \Sigma R^\dagger$,
with $L$ are $R$ being independent $SU(2)$ rotations. Its 
vacuum expectation value, \hbox{$\Sigma_0=\langle \Sigma \rangle$}, 
breaks the chiral 
symmetry down to an $SU(2)$ subgroup. The fluctuations around 
$\Sigma_0$ correspond to the Goldstone-bosons, 
\begin{equation}
  \Sigma = \Sigma_0\,\exp\left\{i\sum_{a=1}^3\pi_a
  \sigma_a/f_\pi\right\} \ .
\label{eq:piondef}
\end{equation}

The mass term explicitly breaks the underlying chiral 
symmetry, and a standard spurion analysis gives, to
second order in $m$, the potential energy\footnote{
Kinetic energy corrections to ${\cal L}_\chi$ need not 
be considered since we are only interested in the vacuum 
state.}
\begin{equation}
\label{vchiral}
  {\cal V}_\chi = 
  - {c_1 \over 4 } {\sf Tr}\left(\Sigma + 
  \Sigma^\dagger\right)
  + {c_2 \over 16 } \left\{
  {\sf Tr}\left(\Sigma + \Sigma^\dagger
  \right)\right\}^2 \,,
\end{equation}
with $c_1\sim m \Lambda^3$ and $c_2\sim m^2 \Lambda^2$. 
Since the Pauli term transforms under chiral 
rotations as does the mass term, and as it is proportional 
to a single power of the lattice spacing, its effects can 
be included (along with those of the mass term) by simply 
making the substitution $m \to m + a \Lambda^2$, where 
$\Lambda$ is required by dimensional analysis. Thus, we 
are led to the potential (\ref{vchiral}) with coefficients\footnote{%
Since we are keeping terms of $O(a^2)$ in $c_2$, we must also include
such terms in ${\cal L}_{\rm eff}$, Eq. (\ref{leff2}). 
This has been done in \cite{ss} and does not change (\ref{c12}).}
\begin{equation}
\label{c12}
  c_1 \sim m\Lambda^3 + a \Lambda^5\,, ~
  c_2 \sim m^2 \Lambda^2 + m a \Lambda^4 + 
  a^2 \Lambda^6 \,.
\end{equation}

We now study the alignment of the condensate, and the
consequent pattern of symmetry breaking, as we reduce $m$ at fixed 
(small) $a$.
For $a\Lambda^2\ll |m| \ll \Lambda$ 
the discretization errors in $c_1$ and $c_2$
are a small correction, and symmetry breaking is as in the continuum.
In particular, since $c_2/c_1\sim |m|/\Lambda\ll 1$, only the first term
in the potential is important, and the condensate aligns as
$\Sigma_0=\pm 1$. Flavor and parity are unbroken.

This situation persists even when $m$ becomes of size $\sim a \Lambda^2$.
Now discretization errors are important, but $c_2/c_1 \sim a \Lambda$ is
still small, and so the condensate is aligned as before.
The effect of the discretization errors is to shift the mass at which
$c_1$ vanishes from $m=0$ to $m'=0$, with $m'= m - a \Lambda^2$
(recall that all constants are being set to unity).
In terms of this  new mass, the coefficients become
\begin{equation}
  c_1 \sim m'\Lambda^3\,,\quad 
  c_2 \sim m'^2 \Lambda^2 + 
  m' a \Lambda^4 + a^2 \Lambda^6 \,.
\label{c12new}
\end{equation}

Finally, when the shifted masses are of $O(a^2)$, {\em i.e.} 
$a m' \sim (a \Lambda)^3$, the coefficients simplify to
\begin{equation}
  c_1 \sim m'\Lambda^3\,,\qquad
  c_2 \sim a^2 \Lambda^6 \,.
\label{c12newest}
\end{equation}
The crucial point is that $c_1\sim c_2$, and, as we show 
below, competition between the two terms can lead to 
spontaneous flavor and parity breaking, 
{\em i.e.} the appearance of an Aoki phase\footnote{When 
the first two terms of the potential are comparable, 
the cubic $c_3$-term is down by a power of $a\Lambda$, 
except for a small region of order $(a\Lambda)^4$ in 
which $c_1 \sim c_2 \sim c_3$.}.
It follows that the width of the Aoki phase is 
$\Delta m_0 \sim a \Delta m' \sim (a \Lambda)^3$.

Let us now examine the potential (\ref{vchiral}) in some 
detail to determine the condensate $\langle \Sigma_0 \rangle$.
Writing \hbox{$\Sigma = A + i
{\bf B} \cdot{\lpmb \sigma}$} with \hbox{$A^2 + 
{\bf B}^2 = 1$}, the potential becomes
\begin{equation}
  {\cal V}_\chi= -c_1 A + c_2 A^2  \ ,
\label{lmint}
\end{equation}
The parameter $A$ is constrained to lie between $-1$ and $+1$ 
inclusive. Under flavor $SU(2)_V$ rotations, $A$ is invariant 
while ${\bf B}$ rotates by an orthogonal transformation. 
Hence, when the vacuum state \hbox{$\Sigma_0 =A_0 + i {\bf B}_0
\cdot{\lpmb \sigma}$} develops a non-zero ${\bf B}_0$, the 
flavor symmetry breaks spontaneously to a $U(1)$ subgroup 
defined by $\exp\left\{i\theta\, \hat{\bf B}_0 \cdot {\lpmb 
\sigma} \right\}$. A non-zero value of ${\bf B}_0$ 
can occur only when $|A_0|$ is strictly less than one. 

\begin{figure}[t]
\epsfxsize=3.0in
\centerline{\epsffile{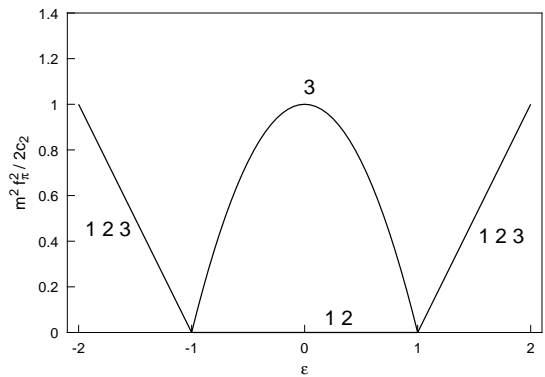}}
\vspace{-0.2truein}
\caption{Pion masses as a function of $\epsilon=c_1/2c_2$ 
for \hbox{$c_2>0$}. The curves are labeled by the flavor 
of the corresponding pion.}
\vspace{-0.2truein}
\label{fig:masseps}
\end{figure}

The sign of $c_2$, which is not predicted by
our analysis, determines two qualitatively different behaviors. 
We discuss $c_2>0$ first, since this gives rise to an Aoki phase.
The potential is then a concave upward parabola 
with absolute minimum at $A_m=c_1/2 c_2 \equiv\epsilon$.
If this minimum lies outside the range $-1$ to $1$, then the vacuum 
is at $A_0= \pm 1$ and flavor symmetry is not spontaneously broken. 
But if $|A_m|<1$, then the vacuum is at $A_0=\epsilon$, and flavor 
symmetry is spontaneously broken. The system thus finds itself in 
the Aoki phase with two massless Goldstone pions. 
The mass spectrum is
\begin{eqnarray}
\label{eq:mpiresultsa}
  m_1^2=m_2^2 = 0 \,, ~
  \frac{m_3^2 f_\pi^2}{2 c_2} = 
  1-\epsilon^2 
   ~{\rm for}~ |\epsilon| \le 1 
  \phantom{\ ,}&&
\\ 
\label{eq:mpiresultsb}
  \frac{m_a^2 f_\pi^2}{2 c_2} = 
  |\epsilon|-1
   ~{\rm for}~ |\epsilon| \ge 1 \ ,&&
\end{eqnarray}
which is illustrated in Fig.~\ref{fig:masseps}.
Recalling from (\ref{c12newest}) that $\epsilon\sim m'/(a^2 \Lambda^3)$,
we see explicitly that the Aoki phase has a width 
$\Delta m_0 \sim a \Delta m' \sim (a\Lambda)^3$.

The $c_2<0$ case gives a concave downward parabola with
an absolute maximum at $A_m=\epsilon$, and therefore
the vacuum is always at $A_0=\pm 1$ for any value of
$\epsilon$. Flavor symmetry is not spontaneously broken,
and there are three degenerate pions with non-zero masses
$ {m_a^2 f^2_\pi}/{2|c_2|} = 1 + |\epsilon|$.

In summary, in the continuum, when the quark mass passes through
zero, the condensate jumps discontinuously from $\Sigma_0=+1$ to
$\Sigma_0=-1$, with the pions becoming massless at the transition
point. On the lattice, the explicit chiral symmetry breaking due
to the Wilson term alters this transition. If $c_2>0$, the
condensate rotates continuously
from $+1$ to $-1$, breaking flavor and parity as it does so. 
If $c_2<0$, the condensate jumps discontinuously as in the continuum,
but the pions never become massless.
This shows how the Aoki phase is intimately related to chiral symmetry
breaking.
Our analysis cannot determine the sign of $c_2$,
and indeed $c_2$ can have different signs for different choices of
discretization of the Dirac operator.
Numerical evidence to date is consistent with $c_2>0$ for both unimproved
and improved Wilson fermions.

We close with a few comments \cite{ss}.
First, one can relate the appearance of the Aoki phase to a non-vanishing
density of zero modes of the hermitian Wilson-Dirac operator.
We stress, however, that our analysis applies only to the unquenched theory,
and has no direct implications for the spectrum of the Wilson-Dirac
operator in the quenched theory.
Second, our analysis applies as well to $O(a)$ improved Wilson fermions.
And, finally, we stress that the phase structure will be different for
$N_f=3$, for which there is no $m\to -m$ symmetry in the continuum.

\end{document}